# Super-Resolution of Radar/Rain Gauge–Analyzed Precipitation Using Gaussian Process Regression with a Steering Kernel

Shoichi AKAMI[1,2], Tsuyoshi T. SEKIYAMA[2], and Mizuo KAJINO[2,3]

[1]*Graduate School of Life and Environmental Sciences, University of Tsukuba, Tsukuba, Japan*

[2]*Meteorological Research Institute, Japan Meteorological Agency, Tsukuba, Japan*

[3]*Faculty of Life and Environmental Sciences, University of Tsukuba, Tsukuba, Japan*

Corresponding author: Shoichi AKAMI, Graduate School of Life and Environmental Sciences, University of Tsukuba, Tennoudai, Tsukuba, Ibaraki 305-8572, Japan. E-mail: s2330311@u.tsukuba.ac.jp

## Abstract

Super-resolution (SR), a technique for estimating a high-resolution (HR) image from a low-resolution image, has been used in meteorology for downscaling and resolution enhancement of observations. SR Gaussian process regression with a steering kernel (SRGP-SK) generates more accurate HR images than the original SR Gaussian process regression (SRGP), but it has not yet been applied to meteorological data. This study applied SRGP-SK to radar/rain gauge–analyzed precipitation for convective and stratiform cases and evaluated the results using the structural similarity index (SSIM) and radially averaged power spectral density (PSD). SRGP-SK achieved SSIM values comparable to those of bicubic interpolation and higher than those of SRGP while reconstructing finer precipitation structures; it recovered variations down to a wavelength of 6 km, compared with 8 km for bicubic interpolation. This difference may correspond to an approximately threefold increase in the number of convective cells. Among the kernel functions compared, the Matérn 5/2 kernel yielded the highest geometric mean PSD ratio. Further investigation of this point could identify the statistical scaling characteristics of the precipitation field. This study evaluates only two precipitation cases; examining more cases is necessary before SRGP-SK can be applied more broadly.

## 1. Introduction

Super-resolution (SR) is a technique for estimating a high-resolution (HR) image from a single low-resolution (LR) image or a set of LR images. In meteorology, SR is used for downscaling and resolution enhancement of observations. Recently, deep learning (DL) has become the mainstream approach to SR (Dong et al. 2015; Vandal et al. 2017; Geiss and Hardin 2020; Sekiyama et al. 2023). DL demonstrates high performance but has three limitations. First, it requires large training datasets. Second, quantifying predictive uncertainty is difficult without advanced techniques, such as Bayesian neural networks or deep ensemble methods (Lakshminarayanan et al. 2017). Third, the model's physical interpretation is challenging (McGovern et al. 2019). In contrast, Gaussian process regression (GPR) offers several advantages: it is highly data efficient when an appropriate kernel function is assumed, provides predictions as probability distributions, enables quantitative uncertainty assessment, and makes underlying assumptions explicit.

He and Siu (2011) proposed SR Gaussian process regression (SRGP), a two-stage method in which the input image is first upsampled and deblurred to obtain an SR image with enhanced edges. However, owing to the complex structure of natural images, simple pixel features and isotropic radial basis function (RBF) kernels cannot accurately capture the relationships among image patches.

To address this, Wang et al. (2021) proposed SRGP with a steering kernel (SRGP-SK), which employs a steering kernel function and an automatic relevance determination (ARD) kernel function. The steering kernel function improves clustering performance, and the ARD kernel function captures common structures among image patches within each cluster, enabling SRGP-SK to generate more accurate SR images than SRGP.

However, SRGP-SK has not yet been applied to meteorology, and its advantages remain unproven. Here, we applied SRGP-SK to radar/rain gauge–analyzed precipitation (RA). Following Geiss and Hardin (2020), we used the radially averaged power spectral density (PSD) of the SR image to verify reconstruction of fine-scale precipitation structures. In addition, we investigated how reconstruction accuracy varies with kernel function. These analyses demonstrate the usefulness of SRGP-SK for meteorological SR and clarify the role of kernel choice in reconstruction.

## 2. Method

### 2.1 Outline of SRGP-SK

SRGP-SK comprises training and test stages. In the training stage, 48 pairs of LR and HR training images $\{L_t, H_t\}_{t=1}^{T}$, obtained every 30 min during the two days preceding the target time, are first acquired (Fig. 1(1)). Here, the running index $t = 1, 2, \dots, T$ denotes the pair index. Subsequently, each $L_t$ is interpolated by a factor of two to yield $\widetilde{L_t}$, whose size equals that of $H_t$. A high-frequency (HF) image $\widetilde{H_t}$ is obtained by subtracting the interpolated image $\widetilde{L_t}$ from $H_t$. Next, $N$ training image patch pairs $\left\{P_n^{(L)}, P_n^{(H)}\right\}_{n=1}^{N}$ are randomly extracted from the image pairs $\{\widetilde{L_t}, \widetilde{H_t}\}_{t=1}^{T}$. For each HF image patch $P_n^{(H)}$ (7 × 7 pixels), its center pixel $p_n^{(H)}$ is extracted to form the training dataset $\mathcal{D} = \{\boldsymbol{x}_n, y_n\}_{n=1}^{N} \triangleq \left\{P_n^{(L)}, p_n^{(H)}\right\}_{n=1}^{N}$, where $\boldsymbol{x}_n$ represents the column-stacked vector of the interpolated LR training image patch $P_n^{(L)}$ and $y_n$ denotes the center pixel of the corresponding HF image patch $P_n^{(H)}$. The steering kernel coefficient (SKC) matrices are then used to represent each $P_n^{(L)}$ and cluster $\mathcal{D}$ into $C$ (= 5) subsets based on Euclidean distance, that is, $\mathcal{D} = \cup_{c=1}^{C} \mathcal{D}_c = \cup_{c=1}^{C} \left\{\boldsymbol{x}_n^{(c)}, y_n^{(c)}\right\}_{n=1}^{N_c}$, with $\sum_{c=1}^{C} N_c = N$ (= 5000) (Fig. 1(2)). Here, the running index $c = 1, 2, \dots, C$ denotes the cluster index, and the running index $n = 1, 2, \dots, N_c$ represents the sample index within subset $\mathcal{D}_c$. Finally, a GPR model is trained for each subset $\mathcal{D}_c$.

In the test stage, the LR test image $S$ is first interpolated by a factor of two to obtain $S_L$, and all image patches of $S_L$ are extracted. As in the training stage, the SKC matrices are used to represent each target image patch and determine the closest cluster $\mathcal{D}_c$. The corresponding GPR model is then used to predict the center pixel of the reconstructed image $S_H$ (Fig. 1(3)). Finally, the center pixel of $S_H$ is added to that of $S_L$ to reconstruct the final SR image $S_S$ through five iterations of back-projection (Fig. 1(4)). The SRGP-SK algorithm is summarized in Fig. 1.

## 2.2 SKC

The steering kernel is an anisotropic kernel adapted to the local structure of an image. Following Wang et al. (2021), the local structure is characterized by applying singular value decomposition (SVD) to the local gradient matrix, from which the covariance matrix $\boldsymbol{R}$ between pixels in the image patch is constructed as follows:

$$\boldsymbol{R} = \xi \boldsymbol{U}_\phi \boldsymbol{E} \boldsymbol{U}_\phi^{\mathrm{T}}, \tag{1}$$

$$\boldsymbol{U}_\phi = \begin{bmatrix} \cos\phi & \sin\phi \\ -\sin\phi & \cos\phi \end{bmatrix}, \tag{2}$$

$$\boldsymbol{E} = \begin{bmatrix} \gamma & 0 \\ 0 & \gamma^{-1} \end{bmatrix}. \tag{3}$$

Here, $\boldsymbol{U}_\phi$ and $\boldsymbol{E}$ denote the rotation matrix and expansion matrix, respectively. Equation (1) is uniquely determined by three parameters: the local structure orientation $\phi$, local structure strength $\gamma$, and shrinkage parameter $\xi$. These parameters are obtained by applying SVD to the local gradient matrix, defined as follows:

$$\boldsymbol{G} = \begin{bmatrix} \vdots & \vdots \\ p_h(i,j) & p_v(i,j) \\ \vdots & \vdots \end{bmatrix}. \tag{4}$$

Here, $\boldsymbol{G}$ is obtained by stacking the horizontal gradient $p_h(i,j)$ and vertical gradient $p_v(i,j)$ of pixel $p(i,j)$ at position $(i,j)$ in the image patch. Applying SVD to $\boldsymbol{G}$ yields

$$\boldsymbol{G} = \boldsymbol{U\Sigma V}^{\mathrm{T}} = \boldsymbol{U}\,\mathrm{diag}([\sigma_1, \sigma_2])\,[\boldsymbol{v}_1, \boldsymbol{v}_2]^{\mathrm{T}}, \tag{5}$$

where $\mathrm{diag}(\cdot)$ denotes a diagonal matrix whose diagonal elements are those of the argument vector, the columns of $\boldsymbol{U}$ are the left singular vectors of $\boldsymbol{G}$, which are not used below, and the singular values satisfy $\sigma_1 \geq \sigma_2 \geq 0$. The orientation $\phi$ is defined by the components of the right singular vector $\boldsymbol{v}_2 = \left[v_2^{(1)}, v_2^{(2)}\right]^{\mathrm{T}}$ corresponding to the smaller singular value $\sigma_2$ of $\boldsymbol{G}$ as follows:

$$\phi = \arctan\left(\frac{v_2^{(1)}}{v_2^{(2)}}\right). \tag{6}$$

The strength $\gamma$ is defined from the singular values $\boldsymbol{\Sigma} = \mathrm{diag}([\sigma_1, \sigma_2])$ of $\boldsymbol{G}$ as follows:

$$\gamma = \frac{\sigma_1 + \alpha_\sigma}{\sigma_2 + \alpha_\sigma}, \tag{7}$$

where the regularization parameter $\alpha_\sigma = 1.0$ improves numerical stability and suppresses pixel-noise effects. The shrinkage parameter $\xi$ is defined as follows:

$$\xi = \left(\frac{\sigma_1\sigma_2 + \alpha_r}{M}\right)^{\frac{1}{2}}, \tag{8}$$

where the regularization parameter $\alpha_r = 0.01$ has the same effect as $\alpha_\sigma$, and $M$ denotes the number of pixels in the image patch.

Once $\boldsymbol{R}$ is computed, the SKC matrix of each image patch is obtained. Let $i_0$ and $j_0$ denote the row and column indices, respectively, of the center pixel $p_n^{(L)}$ in the interpolated LR training image patch $P_n^{(L)}$, and let $\boldsymbol{R}_n^{(L)}$ be the corresponding covariance matrix. The SKC matrix $\boldsymbol{W}_n^{(L)}$ is then defined as follows:

$$\boldsymbol{W}_n^{(L)}(i,j) = \frac{\det\left(\boldsymbol{R}_n^{(L)}\right)}{2\pi\kappa^2} \cdot \exp\left\{-\frac{((i_0,j_0)^{\mathrm{T}} - (i,j)^{\mathrm{T}})^{\mathrm{T}}\boldsymbol{R}_n^{(L)}((i_0,j_0)^{\mathrm{T}} - (i,j)^{\mathrm{T}})}{2\kappa^2}\right\}, \tag{9}$$

where $\boldsymbol{W}_n^{(L)}(i,j)$ denotes the element of $\boldsymbol{W}_n^{(L)}$ in row $i$ and column $j$, $\det(\cdot)$ indicates the determinant of the argument matrix, and the global smoothing parameter is set to $\kappa = 1.6$. Because the SKC matrix numerically represents the local structure of each image patch, we used it as a feature vector for $k$-means clustering.

**2.3 GPR**

In GPR, the center pixels $\boldsymbol{y}^{(c)} = \left\{y_1^{(c)}, y_2^{(c)}, \dots, y_{N_c}^{(c)}\right\}$ corresponding to the training image patches $\boldsymbol{X}^{(c)} = \left\{\boldsymbol{x}_1^{(c)}, \boldsymbol{x}_2^{(c)}, \dots, \boldsymbol{x}_{N_c}^{(c)}\right\}$ are assumed to follow a multivariate Gaussian distribution. As in Wang et al. (2021), this assumption holds when the target image patch is added; that is, the training image patches $\boldsymbol{X} = \left\{\boldsymbol{x}_1^{(c)}, \boldsymbol{x}_2^{(c)}, \dots, \boldsymbol{x}_{N_c}^{(c)}, \boldsymbol{x}^*\right\}$ and the center pixels, including the target pixel, $\boldsymbol{y} = \left\{y_1^{(c)}, y_2^{(c)}, \dots, y_{N_c}^{(c)}, y^*\right\}$ are jointly Gaussian, as follows:

$$\begin{bmatrix} \boldsymbol{y}^{(c)} \\ y^* \end{bmatrix} \sim \mathcal{N}\left(\boldsymbol{0}, \begin{bmatrix} K\left(\boldsymbol{X}^{(c)}, \boldsymbol{X}^{(c)}\right) + \gamma^2 \boldsymbol{I} & K\left(\boldsymbol{X}^{(c)}, \boldsymbol{x}^*\right) \\ K\left(\boldsymbol{x}^*, \boldsymbol{X}^{(c)}\right) & k(\boldsymbol{x}^*, \boldsymbol{x}^*) \end{bmatrix}\right). \tag{10}$$

Here, $\boldsymbol{K} = K\left(\boldsymbol{X}^{(c)}, \boldsymbol{X}^{(c)}\right) + \gamma^2 \boldsymbol{I}$ denotes the covariance matrix of the training image patches, a square matrix of order $N_c$ that includes the noise term $\gamma^2 \boldsymbol{I}$; $\boldsymbol{k}_* = K\left(\boldsymbol{X}^{(c)}, \boldsymbol{x}^*\right)$ represents the covariance (similarity) vector between the target pixel and training image patches; $k_{**} = k(\boldsymbol{x}^*, \boldsymbol{x}^*)$ is the self-covariance (self-similarity) of the target pixel; the superscript $*$ indicates quantities associated with the target pixel; and $\boldsymbol{I}$ denotes the identity matrix of order $N_c$. The posterior distribution of the target pixel $y^*$, given the training dataset $\mathcal{D}_c = \left\{\boldsymbol{x}_n^{(c)}, y_n^{(c)}\right\}_{n=1}^{N_c}$, can then be written as follows:

$$y^* \mid \boldsymbol{X}^{(c)}, \boldsymbol{y}^{(c)}, \boldsymbol{x}^* \sim \mathcal{N}\left(\boldsymbol{k}_*^\top \boldsymbol{K}^{-1} \boldsymbol{y}^{(c)}, k_{**} - \boldsymbol{k}_*^\top \boldsymbol{K}^{-1} \boldsymbol{k}_*\right). \tag{11}$$

For numerical stability, a small jitter term is added to the diagonal elements of covariance matrix $\boldsymbol{K}$ before Cholesky decomposition.

**2.4 Kernel function**

An isotropic kernel, such as the RBF kernel commonly used in GPR, cannot capture complex image structures. Therefore, we adopt the ARD kernel, which incorporates local structural information. With a separate length-scale parameter $\lambda_n$ for the $n$-th pixel, the ARD anisotropic distance is defined as follows:

$$r = \sqrt{\sum_{n=1}^{N_c} \left( \boldsymbol{x}_n^{(c)} - \boldsymbol{x}_n^{(c)\prime} \right)^2 / \lambda_n^2}, \tag{12}$$

where $\boldsymbol{x}_n^{(c)\prime}$ denotes another column-stacked vector in subset $\mathcal{D}_c$.

Building on this, each kernel is expressed as a function of the ARD anisotropic distance $r$. For the exponential kernel,

$$k_{1/2}(r) = \theta_1^{(c)^2} \exp(-r) + \theta_3^{(c)^2} \delta \left( \boldsymbol{x}_n^{(c)} - \boldsymbol{x}_n^{(c)\prime} \right); \tag{13}$$

for the Matérn 3/2 kernel,

$$k_{3/2}(r) = \theta_1^{(c)^2} \left(1 + \sqrt{3} r\right) \exp\left(-\sqrt{3} r\right) + \theta_3^{(c)^2} \delta \left( \boldsymbol{x}_n^{(c)} - \boldsymbol{x}_n^{(c)\prime} \right); \tag{14}$$

for the Matérn 5/2 kernel,

$$k_{5/2}(r) = \theta_1^{(c)^2} (1 + \sqrt{5} r + 5r^2/3) \exp\left(-\sqrt{5} r\right) + \theta_3^{(c)^2} \delta \left( \boldsymbol{x}_n^{(c)} - \boldsymbol{x}_n^{(c)\prime} \right); \tag{15}$$

and for the RBF kernel,

$$k_\infty(r) = \theta_1^{(c)^2} \exp(-r^2/2) + \theta_3^{(c)^2} \delta \left( \boldsymbol{x}_n^{(c)} - \boldsymbol{x}_n^{(c)\prime} \right). \tag{16}$$

These kernels differ only in terms of the smoothness imposed on the function. Here, $\theta_1^{(c)^2}$ denotes the signal variance, $\theta_3^{(c)^2}$ is the noise variance, and $\delta$ represents the Kronecker delta. The length-scale parameter $\lambda_n$ acts as a relevance parameter: a smaller $\lambda_n$ makes the kernel more sensitive to the $n$-th pixel, assigning greater importance to it. Notably, the hyperparameter values differ among subsets.

For each subset $\mathcal{D}_c$, $\theta_1^{(c)}$ and $\theta_3^{(c)}$ denote the standard deviations of the signal and the noise, respectively, and are initialized as follows:

$$\theta_1^{(c)} = \mathrm{std}\left(\left\{\boldsymbol{x}_n^{(c)}\right\}_{n=1}^{N_c}\right), \tag{17}$$

$$\theta_3^{(c)} = \mathrm{std}\left(\left\{y_n^{(c)}\right\}_{n=1}^{N_c}\right), \tag{18}$$

where $\mathrm{std}(\cdot)$ denotes the standard deviation of the argument data, and $\left\{\boldsymbol{x}_n^{(c)}\right\}_{n=1}^{N_c}$ and $\left\{y_n^{(c)}\right\}_{n=1}^{N_c}$ represent the training image patches and corresponding center pixels in subset $\mathcal{D}_c$ (Wang et al. 2021).

The length-scale parameter $\lambda_n^{(c)}$ is initialized from the average SKC matrix $\overline{\boldsymbol{W}_c}$, obtained by averaging the normalized SKC matrices of all image patches in subset $\mathcal{D}_c$, as follows:

$$\lambda_n^{(c)} = \sqrt{\frac{1}{\left[\overline{\boldsymbol{W}_c}\right]_{nn}}}, n = 1,2,\dots,N_c, \tag{19}$$

where $\left[\overline{\boldsymbol{W}_c}\right]_{nn}$ denotes the $n$-th diagonal component of $\overline{\boldsymbol{W}_c}$. All hyperparameters $\theta_1^{(c)}$, $\theta_3^{(c)}$, and $\lambda_n^{(c)}$ are then optimized for each subset $\mathcal{D}_c$ by maximizing the marginal likelihood.

## 2.5 PSD

This study used PSD to compare the reconstruction accuracy of bicubic interpolation, SRGP, and SRGP-SK. PSD indicates how well each method reconstructs fine-scale precipitation structures and reveals artifacts. The PSD is defined as follows:

$$\mathrm{PSD} = \overline{\left|F_2\{\widehat{S_s}\}\right|^2}, \tag{20}$$

where the overbar denotes radial averaging in Fourier space, $F_2\{\cdot\}$ represents the two-dimensional Fourier transform, and $\widehat{S_s}$ is the SR image generated by each method (Geiss and Hardin 2020).

## 3. Data and Target Region

We used the RA product provided by the Japan Meteorological Agency (JMA). RA is a 1-km-grid, 1-h accumulated precipitation product produced by merging radar reflectivity with rain gauge observations from networks operated by JMA (including the Automated Meteorological Data Acquisition System), the Ministry of Land, Infrastructure, Transport, and Tourism, and local governments (Makihara 1996; Nagata 2011). The product is available every 30 min. This study focused on two representative cases: convective precipitation (CONV) over Hiroshima Prefecture at 21:00 UTC on July 6, 2018 and stratiform precipitation (STRA) over Okinawa Prefecture at 00:00 UTC on June 2, 2011. We selected the CONV case because it produced record-breaking heavy rainfall (Tsuguti et al. 2019) and the STRA case because it exhibited a high rainfall rate for stratiform precipitation (Oue et al. 2015). The target regions cover 32.7501–35.2499°N and 130.6250–134.3750°E for CONV and 26.2501–28.7499°N and 126.1250–129.8750°E for STRA (Fig. 2).

## 4. Results

We evaluated the SR images objectively using the structural similarity index measure (SSIM) and subjectively by visual inspection. For CONV, SRGP-SK achieved reconstruction accuracy comparable to that of bicubic interpolation in terms of SSIM. The SR image reconstructed by SRGP-SK also showed high sharpness and fine-scale precipitation structures, particularly in the region enclosed by the black frame (Fig. 3).

For STRA, SRGP-SK achieved reconstruction accuracy comparable to that of bicubic interpolation in terms of SSIM. The SR image reconstructed by SRGP-SK also exhibited high sharpness and fine-scale precipitation structures, as in CONV (Fig. 4).

We compared the radially averaged PSD [Eq. (20)] to evaluate SR images in Fourier space. For CONV, SRGP-SK preserved more power than the other methods at wavelengths shorter than the LR Nyquist wavelength of the interpolated LR training image, corresponding to fine-scale precipitation structures. Moreover, no features indicative of artifacts were observed. Notably, the PSD curve for SRGP-SK nearly overlapped that of the ground truth up to the LR Nyquist wavelength, whereas the other methods lost power even at long wavelengths (Fig. 5a). For STRA, features similar to those in CONV were observed (Fig. 5b). Notably, SRGP-SK reconstructed fine-scale precipitation structures down to a wavelength of 6 km, whereas bicubic interpolation reconstructed them only down to a wavelength of 8 km for both cases.

Table 2 lists the objective measures (SSIM and geometric mean PSD ratio) used to evaluate SRGP-SK with different kernel functions across the two cases. For CONV, the exponential kernel performed best on SSIM, whereas the Matérn 3/2 kernel performed best on the geometric mean PSD ratio. For STRA, the exponential kernel again performed best on SSIM, whereas the Matérn 5/2 kernel performed best on the geometric mean PSD ratio. The optimal kernel differed between measures because SSIM assesses whether identical structures appear at the same location as in the ground truth, whereas the geometric mean PSD ratio assesses whether each scale contains the same amount of precipitation structure as the ground truth. Accordingly, a high SSIM but low geometric mean PSD ratio indicates that overall locations are reconstructed but fine-scale precipitation structures are underestimated. Conversely, a low SSIM but high geometric mean PSD ratio indicates that plausible textures are placed at incorrect locations. Furthermore, all kernel functions yielded comparable geometric mean PSD ratios for CONV, whereas marked differences emerged for STRA.

## 5. Discussion

SRGP-SK generated accurate SR images for CONV and STRA (Figs. 3 and 4) because it employs the steering kernel to cluster image patches and the ARD kernel, which assigns an independent length-scale parameter to each input variable. In contrast, SRGP produced less accurate SR images because it cannot fully exploit prior information in natural images (Xie et al. 2013). Moreover, the SSIM values of bicubic interpolation and SRGP-SK were comparable because bicubic interpolation was used for downsampling and upsampling in this study.

SRGP-SK exhibited a higher PSD than the other methods at wavelengths shorter than the LR Nyquist wavelength (Fig. 5). This result can be explained as follows: First, because bicubic interpolation does not reconstruct HF components absent from the LR image, it exhibited low PSD at wavelengths shorter than the LR Nyquist wavelength. Second, because SRGP fits a single GPR jointly to image patches with different structures, its length-scale parameter is optimized to reflect the average smoothness of the entire dataset. Consequently, the predicted HF image power decreased, and SRGP exhibited a relatively low PSD at wavelengths shorter than the LR Nyquist wavelength. In contrast, SRGP-SK trains GPR models separately for clusters of similar image patches, allowing the length-scale parameter to adapt to each cluster's local structure; accordingly, SRGP-SK exhibited high PSD at wavelengths shorter than the LR Nyquist wavelength.

Moreover, because the steering kernel suppresses smoothing across edges, the PSD at wavelengths longer than the LR Nyquist wavelength nearly matched that of the ground truth.

Bicubic interpolation reconstructed precipitation structures down to a wavelength of 8 km, whereas SRGP-SK reconstructed them down to a wavelength of 6 km. As area scales with the square of wavelength, this implies that the smallest identifiable precipitation structure area is reduced to approximately half. This difference in the shortest resolvable wavelength has direct implications for representing convection. Although the significance of this 2-km difference is challenging to interpret, we provide an illustrative example. Miyamoto et al. (2013) performed global subkilometer simulations and examined the dependence of the number of convective cells on grid spacing. Although the model grid spacing and resolvable wavelength presented here are not directly comparable, the difference between 8 km and 6 km corresponds to an approximately threefold increase in the number of convective cells, from $7.54 \times 10^3$ to $2.38 \times 10^4$. Currently, no firm evidence indicates that the 2-km improvement in the resolvable wavelength offered by SRGP-SK is practically advantageous, but its potential importance cannot be ruled out.

For STRA, the geometric mean PSD ratio differed markedly among the kernel functions (Table 2), attributable to the sparsity of precipitation structures at wavelengths shorter than the LR Nyquist wavelength. Because such precipitation structures are almost absent from the LR image, the SR image is weakly constrained by the LR image in this band, and the PSD is dominated by each algorithm’s intrinsic extrapolation behavior. Consequently, differences among kernel functions are reflected primarily in the PSD within this band. Among the tested kernels, the Matérn 5/2 kernel yielded the highest geometric mean PSD ratio for STRA. This result suggests that kernel smoothness affects the reconstructed spectrum. However, the kernel function in SRGP-SK defines the covariance between input image patches, thereby encoding the prior assumption that similar interpolated LR image patches yield similar center pixels of the corresponding HF image patches. Therefore, its interpretation in terms of the precipitation field remains difficult.

## 6. Conclusion

We applied SRGP-SK to RA and showed that it achieved higher reconstruction accuracy than SRGP. SRGP-SK also reconstructed precipitation structures down to shorter wavelengths than bicubic interpolation. Assessing the practical value of this insight will require evaluating the number of convective cells in the SR images, which is left for future work.

We further showed that the Matérn 5/2 kernel yielded the highest geometric mean PSD ratio. Realizing the potential of this result requires a framework in which the kernel function is defined in physical space, as in Wang et al. (2015). In such a framework, the spectral density of the kernel function corresponds analytically to the precipitation field spectrum, substantially improving kernel interpretability. Owhadi (2023) demonstrated that the residual obtained by substituting an approximate solution of Gaussian process hydrodynamics into the vorticity equation decreases markedly when the kernel function is consistent with the power law of the velocity increment. Conversely, this implies that the statistical scaling characteristics of the precipitation field could be identified by estimating the kernel function that minimizes the error. However, this approach requires the kernel function to be defined in physical space and characterizes only second-order statistics. Finally, examining more cases is necessary before SRGP-SK can be applied more broadly.

## Declarations

### Acknowledgements

The authors used Claude (Anthropic) and Enago for English language editing of the manuscript. The authors reviewed all output of Claude and take full responsibility for the content of this publication.

### Funding

This research was supported by JST SPRING (Grant Number JPMJSP2124). This research was also supported by the Environmental Research and Technology Development Fund (Grant Number JPMEERF20252M03) of the Environmental Restoration and Conservation Agency of Japan (ERCA), and by a Grant-in-Aid for Scientific Research (KAKENHI) (Grant Numbers JP23K21747 and JP25H00784) from the Japan Society for the Promotion of Science.

### Competing Interests

The authors declare that the authors have no competing interests as defined by Springer, or other interests that might be perceived to influence the results and/or discussion reported in this paper.

### Author Contributions

SA: Conceptualization (lead); Data curation (lead); Formal analysis (lead); Investigation (lead); Methodology (lead); Project administration (lead); Resources (supporting); Software (lead); Supervision (supporting); Validation (lead); Visualization (lead); Writing - original draft (lead); Writing - review and editing (lead); Funding acquisition (supporting). TTS: Data curation (supporting); Formal analysis (supporting); Methodology (supporting); Writing - review and editing (supporting); Funding acquisition (supporting). MK: Investigation (supporting); Data curation (supporting); Project administration (supporting); Resources (lead); Supervision (lead); Writing - review and editing (supporting); Funding acquisition (lead).

**Data Availability**

The data and the source code that support the findings of this study are available from the corresponding author upon reasonable request.

## List of Figure Captions

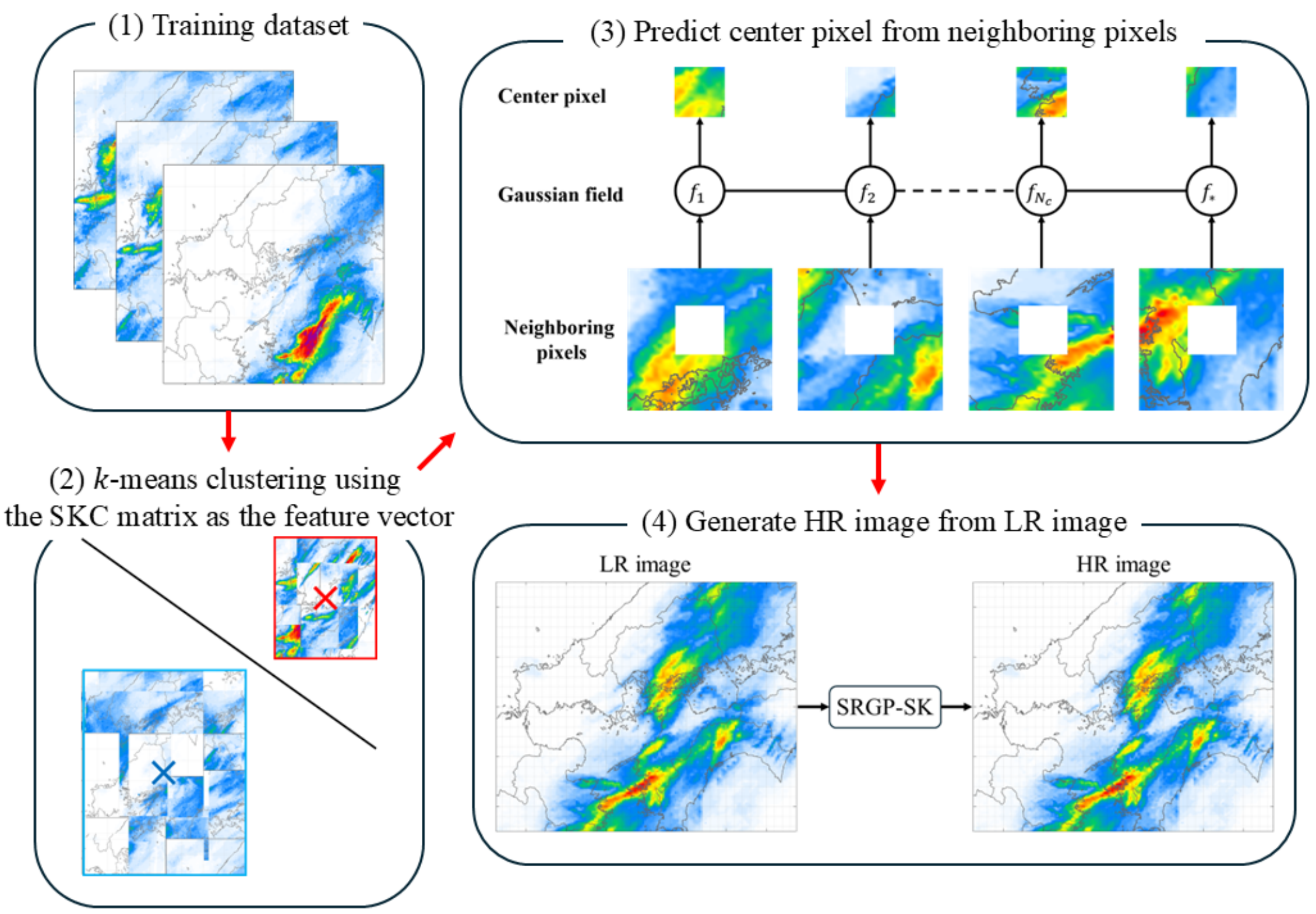

**Fig. 1.** Flowchart of the SR procedure based on SRGP-SK: (1) RA images at times other than the target time are collected to construct the training dataset. (2) The training dataset is clustered by $k$-means using the SKC matrix as the feature vector, yielding GPR models adapted to local image structures. The red and blue crosses denote cluster centers; the red cluster comprises center pixels of HF image patches, the blue cluster comprises interpolated LR target image patches, and the diagonal line represents the decision boundary. (3) Each target image patch is assigned to a cluster, and the center pixel of the corresponding HF image patch is predicted from the collocated interpolated LR target image patch using the GPR model of that cluster. The inset shows the graphical model: squares denote pixels of the interpolated LR target image patch; circles denote the unknown Gaussian field; the horizontal line

represents a set of fully connected nodes; $f$ denotes a function following a Gaussian process; the subscript $n = 1, 2, \ldots, N_c$ represents the sample index; and the superscript $*$ denotes quantities associated with the target pixel. (4) The HR image is generated from the LR image using SRGP-SK.

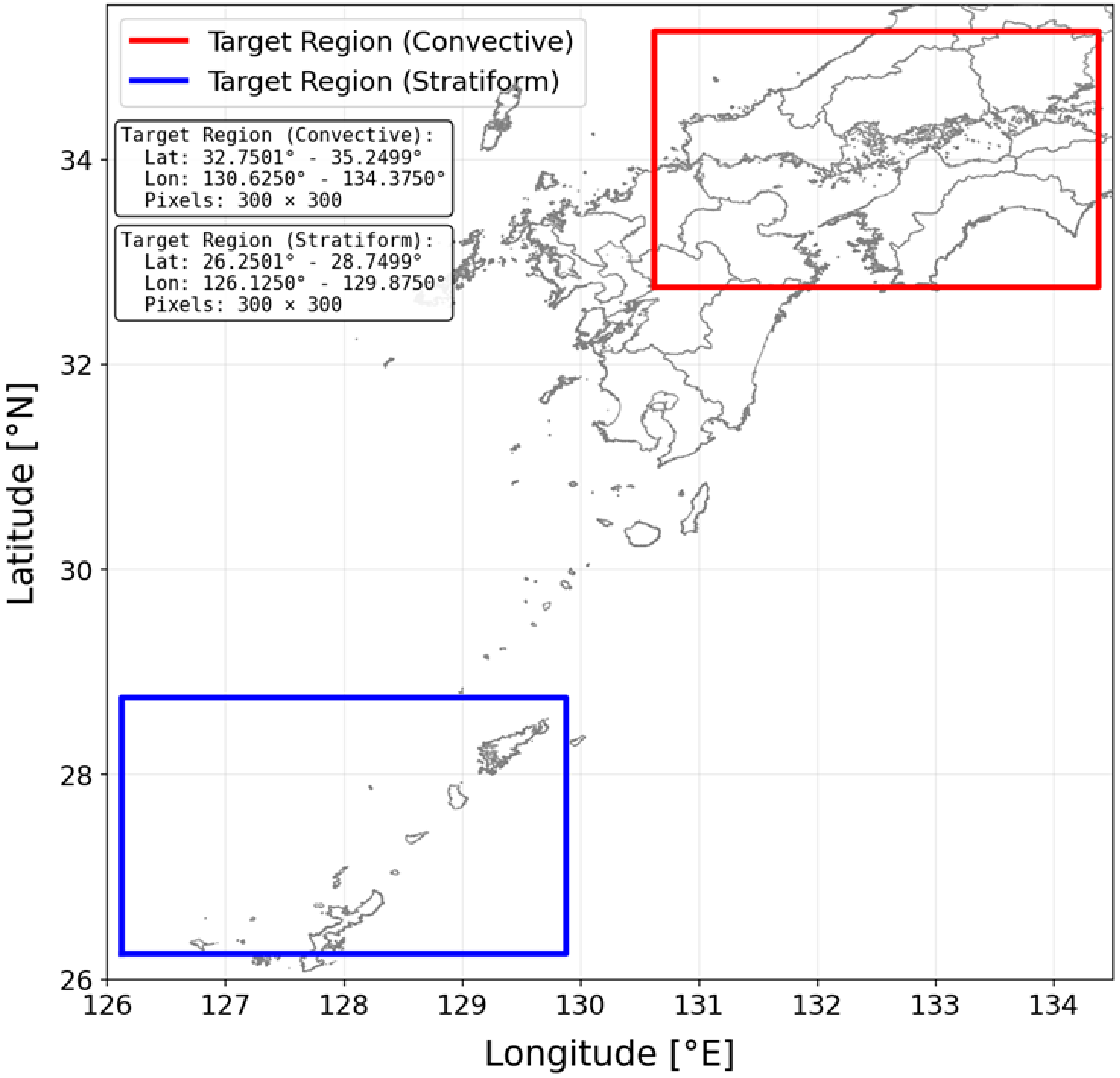

**Fig. 2.** Target regions of this study. The red square denotes the target region for the CONV case (over Hiroshima Prefecture at 21:00 UTC on July 6, 2018), and the blue square denotes that for the STRA case (over Okinawa Prefecture at 00:00 UTC on June 2,

2011). The CONV case region covers 32.7501–35.2499°N and 130.6250–134.3750°E, and the STRA case region covers 26.2501–28.7499°N and 126.1250–129.8750°E. The RA grid spacing is 0.012500° in longitude and 0.008333° in latitude, resulting in nonrounded coordinate values. Both regions comprise 300 × 300 pixels, forming square images.

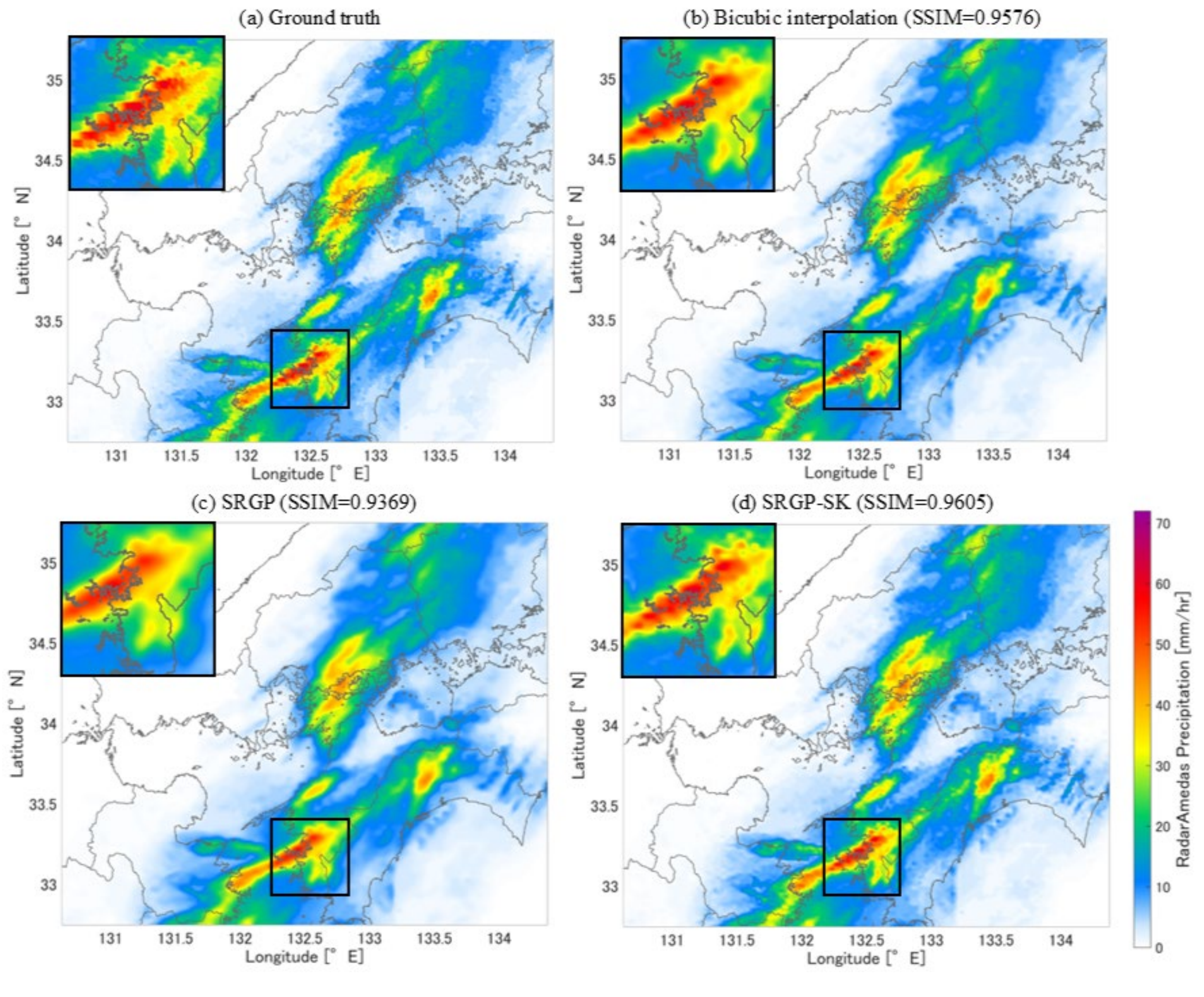

**Fig. 3.** Comparison of 2× SR results for the CONV case: (a) ground truth, (b) bicubic interpolation, (c) SRGP, and (d) SRGP-SK. The SSIM with respect to the ground truth is shown above each reconstruction (panels b–d). The inset in the upper-left corner of each panel shows a magnified view of the region enclosed by the black frame.

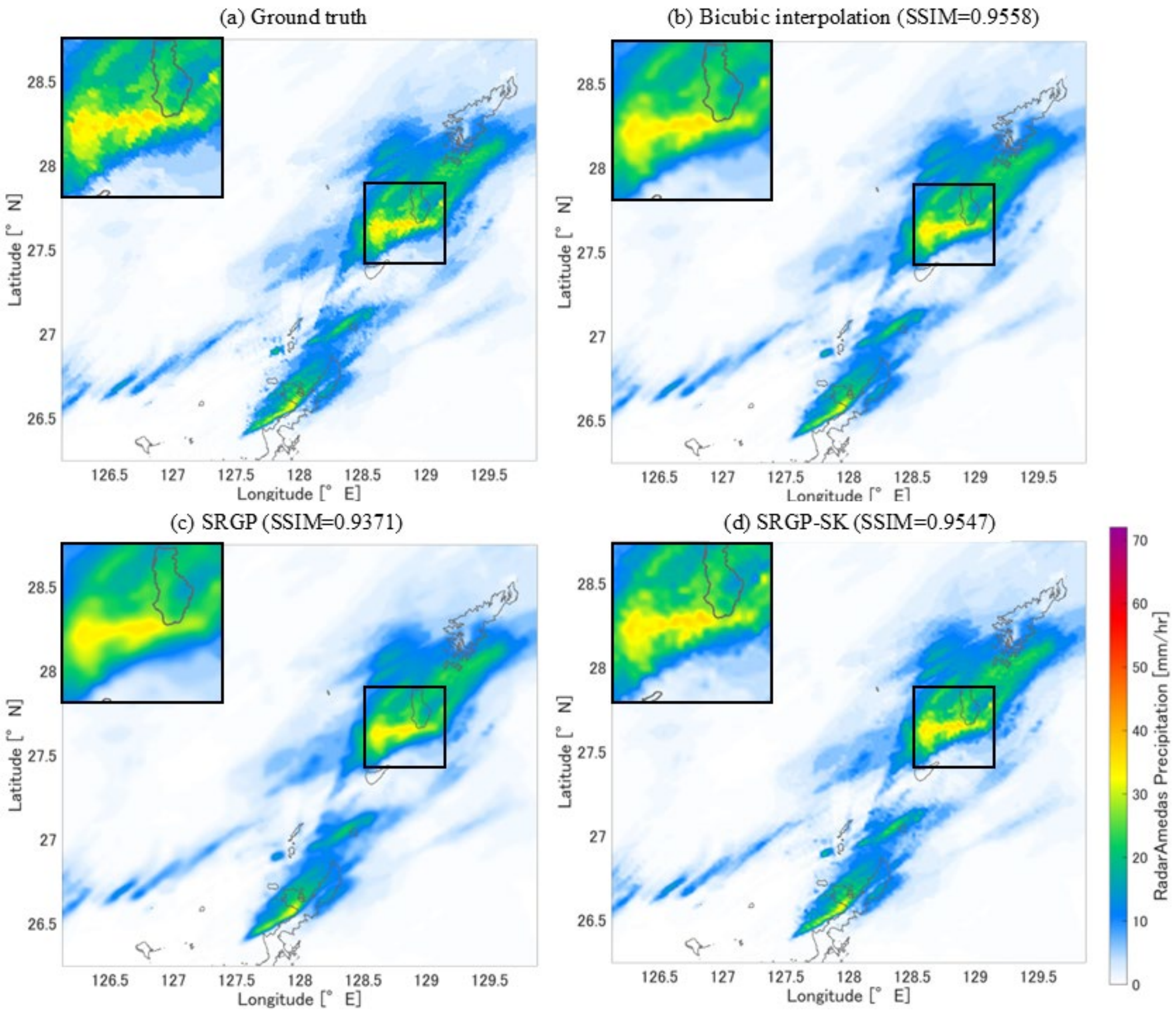

**Fig. 4.** Comparison of 2× SR results for the STRA case: (a) ground truth, (b) bicubic interpolation, (c) SRGP, and (d) SRGP-SK. The SSIM with respect to the ground truth is shown above each reconstruction (panels b–d). The inset at the upper-left corner of each panel shows a magnified view of the region enclosed by the black frame.

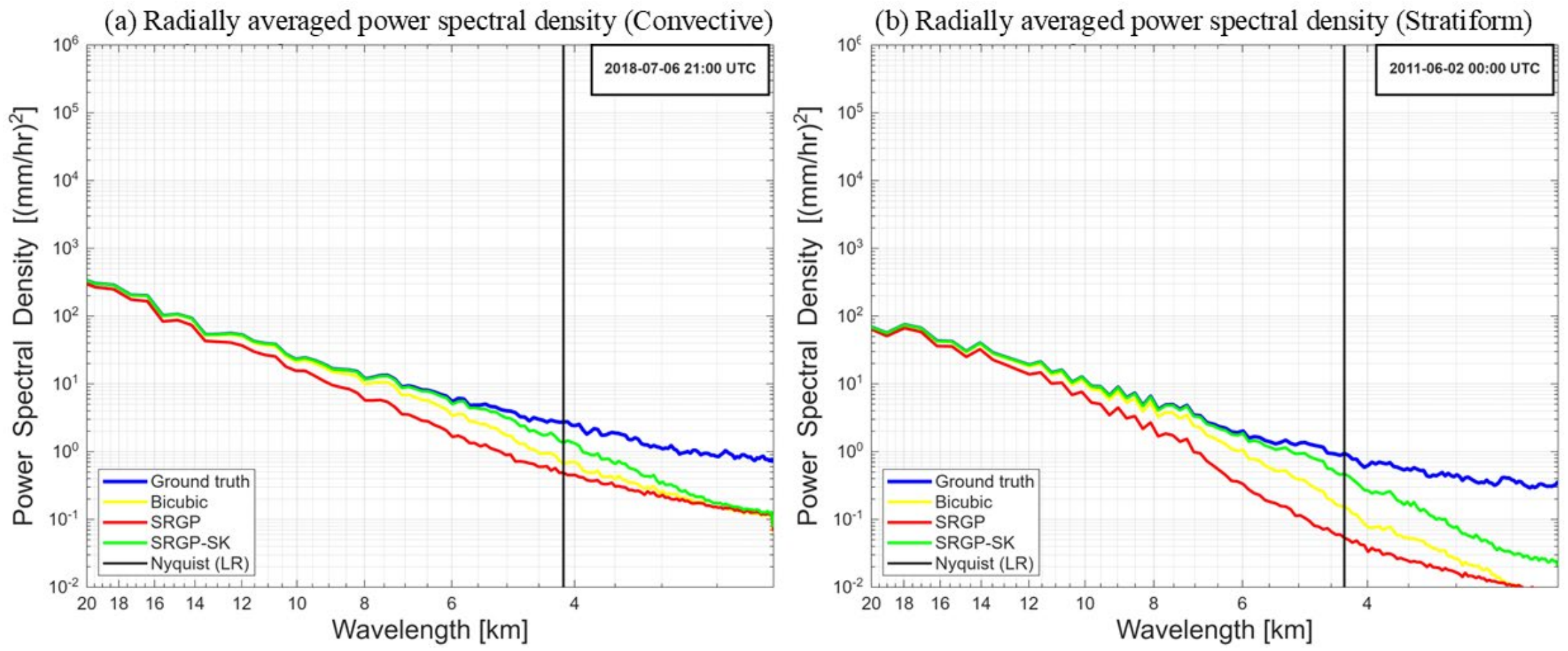

**Fig. 5.** Radially averaged PSD of the ground truth, bicubic interpolation, SRGP, and SRGP-SK: (a) PSD for the CONV case (over Hiroshima Prefecture at 21:00 UTC on July 6, 2018) and (b) PSD for the STRA case (over Okinawa Prefecture at 00:00 UTC on June 2, 2011). The vertical axis represents PSD [(mm/hr)$^2$], and the horizontal axis represents wavelength [km]. Both axes are logarithmic, and the horizontal axis is reversed. The ground truth comprises 300 × 300 pixels, and the vertical black line marks the Nyquist wavelength of the interpolated LR training image (150 × 150 pixels) input to SRGP-SK: 4.15 km for the CONV case and 4.31 km for the STRA case. The blue, yellow, red, and green lines indicate the PSD of the ground truth, bicubic interpolation, SRGP, and SRGP-SK, respectively. The higher PSD of SRGP-SK (green line) at wavelengths shorter than the LR Nyquist wavelength indicates that SRGP-SK preserves fine-scale precipitation structures much better than the other methods (yellow and red lines). SRGP-SK also preserves nearly all power at wavelengths longer than the LR Nyquist wavelength (the region between the vertical axis and the black line), remaining closest to the ground truth among the reconstructions.

**List of Table Captions**

Table 1. Algorithm of the SRGP-SK method.

**Input:** LR/HR training image pairs $\{L_t, H_t\}_{t=1}^{T}$, SR magnification factor $s$, and LR test image $S$

**Output:** SR image $S_s$ generated by SRGP-SK

**Training stage:**

1. Apply bicubic interpolation to each LR training image $L_t$ to generate an interpolated LR training image $\widetilde{L_t}$ with the same size as the HR training image $H_t$.

2. Generate an HF image $\widetilde{H_t}$ by subtracting $\widetilde{L_t}$ from $H_t$.

3. Randomly sample $N$ training image patch pairs $\left\{P_n^{(L)}, P_n^{(H)}\right\}_{n=1}^{N}$ from $\left\{\widetilde{L_t}, \widetilde{H_t}\right\}_{t=1}^{T}$.

4. Generate training dataset $\mathcal{D} = \{\boldsymbol{x}_n, y_n\}_{n=1}^{N} \triangleq \left\{P_n^{(L)}, p_n^{(H)}\right\}_{n=1}^{N}$ by extracting the center pixel $p_n^{(H)}$ from each HF image patch $P_n^{(H)}$.

5. Compute the SKC matrix for each interpolated LR training image patch $P_n^{(L)}$ and cluster the training dataset $\mathcal{D}$ into $\mathcal{D} = \cup_{c=1}^{C} \mathcal{D}_c = \cup_{c=1}^{C} \left\{\boldsymbol{x}_n^{(c)}, y_n^{(c)}\right\}_{n=1}^{N_c}$. As the sample number $N_c$ in subsets $c = 1, 2, \dots, C$ sums to the total sample number $N$, $\sum_{c=1}^{C} N_c = N$ holds.

6. Initialize the hyperparameters for each subset $\mathcal{D}_c$ following Eqs. (17)–(19).

7. Train a GPR model for each subset $\mathcal{D}_c$.

**Test stage:**

8. Apply bicubic interpolation to the LR test image $S$ with SR magnification factor $s$ to generate an interpolated LR test image $S_L$.

9. Initialize the reconstructed image as $S_H = S_L$.

10. Extract an interpolated LR target image patch $\boldsymbol{x}^*$ from $S_L$.

11. Search for the nearest neighbor subset $\mathcal{D}_c$ based on the SKC matrix of $\boldsymbol{x}^*$.

12. Predict the target pixel $y^*$ using the GPR model of that subset.

13. Add the center pixel of the interpolated LR target image patch $\mathrm{cen}(\boldsymbol{x}^*)$ to the target pixel $y^*$ to obtain the SR pixel $s^*$, where $\mathrm{cen}(\cdot)$ denotes the center pixel of its argument image patch.

14. Replace the corresponding pixel in $S_H$ with $s^*$.

15. Apply iterative back-projection to $S_H$ to obtain the SR image $S_s$.

Table 2. SSIM and geometric mean PSD ratio [%] for different kernel functions across the two precipitation cases. Exp, Matérn3, Matérn5, and RBF denote the exponential, Matérn 3/2, Matérn 5/2, and RBF kernels, respectively. The geometric mean PSD ratio is obtained by calculating the ratio of the geometric mean PSD values of SRGP-SK and the ground truth at wavelengths shorter than the LR Nyquist wavelength. The best score in each row is highlighted in bold.

| Case | SSIM | | | |
|---|---|---|---|---|
| | Exp | Matérn3 | Matérn5 | RBF |
| CONV | **0.9607** | 0.9583 | 0.9584 | 0.9605 |
| STRA | **0.9594** | 0.9545 | 0.9475 | 0.9547 |
| Case | Geometric mean PSD ratio [%] | | | |

| | Exp | Matérn3 | Matérn5 | RBF |
|---|---|---|---|---|
| CONV | 28.28 | **28.84** | 27.85 | 25.86 |
| STRA | 13.36 | 25.08 | **39.34** | 16.27 |